\documentclass[showpacs,preprintnumbers,amsmath,amssymb,preprint]{revtex4}
\usepackage{graphics}
\begin{document}

\title{Multifragmentation and the symmetry term of the nuclear equation of
state}

\author{Ad. R. Raduta$^{1}$, F. Gulminelli$^{2}$}

\address{$^{1}$NIPNE, Bucharest-Magurele, POB-MG 6, Romania\\
$^{2}$LPC (IN2P3-CNRS/Ensicaen et Universite), F-14050 Caen cedex, France}

\begin{abstract}
We investigate the possibility to extract the symmetry energy from multifragmentation data.
The applicability of the grandcanonical formula earlier proposed by Ono {\it et al.} 
[Phys. Rev. C {\bf 68}, 051601(R)] 
in the case of finite excited nuclei is tested within a microcanonical framework.
%Accurate results are obtained only for relatively light clusters emitted by highly excited
%large nuclei.
Relatively good results are obtained except for large residual nuclei,
especially when large sources are highly excited.
Effects of secondary particle emission and the extent in which
relevant information may be inferred from experimental observables are finally discussed.
\end{abstract}
\maketitle

\section{Introduction}

Isospin dependent phenomena are attracting increasing interest as they hold the promise
of revealing the asymmetry term of the nuclear equation of state.
At normal nuclear density this quantity dictates the structure of the neutron-rich and
neutron deficient isotopes, %\cite{nuclstruct}, 
while in other domains of density and temperature
its behavior is reflected in a variety of astrophysical phenomena, such as
the structure and evolution of neutron stars and the dynamics of supernovae explosions. 
%\cite{astro}.
In nuclear multifragmentation reactions, the asymmetry term influences
the neutron-proton composition of the break-up fragments.

Interpreting multifragmentation in the light of first-order phase transitions
in multicomponent systems,
the neutron enrichment of the gas phase with respect to the liquid phase
comes out as a natural consequence of Gibbs equilibrium criteria
and a connection between phases chemical composition and the symmetry term
can be established \cite{mueller,gulminelli}.
Interesting enough, the phenomenon of isospin fractionation which is 
systematically observed in analyses of multifragmentation 
data \cite{xu,geraci,martin,shetty,botvina}, seems to be a generic feature of phase
separation independent of the equilibrium Gibbs construction \cite{isospinfrac}. 
Indeed, dynamical models of heavy ion collisions \cite{baoanli,sil,lee,ono,ditoro}
where fragment formation is essentially ruled by the out of equilibrium process of 
spinodal decomposition also exhibit fractionation.
Adopting an equilibrium scenario for the break-up stage of a multifragmenting system,
Ono {\it et al.} \cite{ono} derive an approximate grandcanonical  expression which connects
the symmetry term with the isotopic composition of 
fragments obtained in the break-up stage of two 
sources with similar sizes in identical thermodynamical states 
and differing in their isospin content,
\begin{equation}
C_{sym}=-\frac{\alpha_{12}T}{4 \left[ \left( \frac{Z_1}{A_1}\right)^2-
\left( \frac{Z_2}{A_2}\right)^2\right] },
\label{eq:csym_ono}
\end{equation}
under the hypothesis that the isotopic distributions are essentially Gaussian
and that the free energies contain only bulk terms.
Here, $\alpha_{12}$ is the isoscaling slope parameter,
$Z_i/A_i$ stands for the isospin asymmetry of a fragment produced
by the source $i(=1,2)$ and $T$ is the temperature of the decaying systems. 

In the limit of vanishing temperature, fractionation can be neglected
and $Z_i/A_i$ can be replaced by the corresponding quantity of the sources \cite{botvina}
giving
\begin{equation}
C_{sym}=-\frac{\alpha_{12}T}{4 \left[ \left( \frac ZA\right)_{s1}^2-
\left( \frac ZA\right)_{s2}^2\right] }.
\label{eq:csym_tsang}
\end{equation}
This expression, which was first proposed in the framework of the EES model \cite{tsang},
as well as Eq. (\ref{eq:csym_ono}),
has been used extensively on experimental data and results ranging from
values compatible with the ground state bulk symmetry term
($\approx$ 25 MeV) to about half this value have been interpreted as sign of dilute matter at
freeze-out.
Thus, $C_{sym}$=14 MeV has been obtained in FRS experiments using 
mid-peripheral collisions of 1 GeV/nucleon $^{124,136}$Xe beams on Pb targets \cite{frs};
values decreasing from 25 to 15 MeV have resulted from a
variety of heavy ion collisions induced by few tens MeV/nucleon projectiles
studied by the Texas A\&M group \cite{souliotis03,shetty04,souliotis06,iglio},
while from the fragmentation of excited target residues produced in the peripheral collisions
of $^{12}$C on $^{112,124}$Sn at 300 and 600 MeV/nucleon incident energies
the INDRA-ALADIN collaboration extracts a symmetry coefficient
which decreases from 25 MeV for peripheral collisions to 15 MeV for central collisions
\cite{lefevre}. 

If these values really correspond to the break-up stage of the decay, the implications are
dramatic. First, if the primary fragments at the equilibrated freeze-out are diluted,
statistical models which described successfully a wealth of experimental data
over more than two decades should be completely reformulated.
Starting from the geometrical extension of fragments which dictates the free volume,
to the binding and internal excitation energies which enter energy balance and
affect fragments partitions, all quantities need serious reconsideration.
If, on the other hand,
this diluteness would reflect inter-fragment interactions at break-up,
as advanced by Botvina {\it et al.} \cite{botvina},
the situation would be even more severe as it would refute the fundamental hypothesis
of statistical models, namely the lack of any fragment interaction except the Coulomb one.

Given the implications of such measurements,
the present work aims to investigate the possibility to extract the symmetry
term from multifragmentation data using Eq. (\ref{eq:csym_ono}).
Sticking to the equilibrium hypothesis, Ono's formula has three drawbacks:
a) it is grandcanonical while the grandcanonical approximation is known to be
acceptable only for relatively light fragments emitted by
large systems with high excitation energies,
b) it does not account for full mass dependence of the binding energy
%it neglects surface energy contribution
and c) it holds
for the break-up stage of the reaction, impossible to access experimentally.
Eq. (\ref{eq:csym_tsang}), which was employed to obtain $C_{sym}$ values as low
as $\approx$ 10 MeV \cite{lefevre}, contains the additional approximation of 
neglecting isospin fractionation.

A quantitative estimation of the possible deformations induced by these 
effects can be done in the framework of a microcanonical equilibrium model \cite{micromodels}. 
The less serious objection against Eq. (\ref{eq:csym_ono}),
possible deviations produced by omitting the
contributions of surface \cite{danielewicz}, Coulomb and asymmetry terms,
may be overcome taking into account the full dependence
of the binding energy on the system size and will be addressed first.

Thus, adopting the grandcanonical expression of the isotopic yield of an emitted cluster
with $N$ neutrons and $Z$ protons,
\begin{equation}
Y(N,Z) \propto \exp \left[ \frac 1T \left( B(N,Z)+\mu_n N+\mu_p Z\right)\right],
\label{granca}
\end{equation}
where $\mu_n$ and $\mu_p$ stand for neutron and proton chemical potentials and
$B(N,Z)$ represents the binding energy,
we use as in Ref. \cite{ono} the Gaussian shape of $Y(N,Z)$ distributions as a function of
$N$ ($Z$) when $Z$ ($N$) is kept fixed to approximate
the value of $N$ ($Z$)
corresponding to the maximum of the distribution with its average value.
Under these assumptions, the equation
\begin{eqnarray}
\nonumber
\frac{\partial \left( \ln Y(N,Z)\right)}{\partial N}&=&0\\
&=&\frac1T \left[ \frac{\partial B(N,Z)}{\partial N}|_Z+\mu_n \right],
\end{eqnarray}
applied for two isospin different similar nuclei in identical equilibrium states,
relates the isoscaling slope parameter $\alpha$ with the difference of
partial derivatives of $B(A,Z)$ as a function of $N$.

If we consider for the binding energy a liquid drop parameterization 
including surface and Coulomb terms,
\begin{equation}
B(A,Z)=(a_v A - a_s A^{2/3})-(a_i a_v A - a_i a_s A^{2/3}) I^2 + a_c Z^2/A^{1/3} +
a_a Z^2/A,
\label{eq:bld}
\end{equation}
and account for full mass dependence of the bulk+surface,
isospin-dependent and Coulomb contributions, the resulting equation
will acquire additional terms corresponding to surface, Coulomb and asymmetry energies,
\begin{eqnarray}
\nonumber
C_{sym}&=&\frac{-\alpha_{12} T+\frac23 a_s \left( <A_1>^{-1/3} - <A_2>^{-1/3}\right)}
{4\left[ \left(\frac{Z}{<A_1>}\right)^2 -  \left(\frac{Z}{<A_2>}\right)^2 \right]}\\
\nonumber
&+&\frac{\frac13 Z^2 a_c \left( <A_1>^{-4/3} - <A_2>^{-4/3}\right)
+ a_a Z^2 \left( <A_1>^{-2} - <A_2>^{-2}\right)}
{4\left[ \left(\frac{Z}{<A_1>}\right)^2 -  \left(\frac{Z}{<A_2>}\right)^2 \right]} \\
&+&\frac 13 \frac{a_i a_s \left( <A_1^{-1/3}><I_1>^2-<A_2^{-1/3}><I_2>^2 \right)}
{4\left[ \left(\frac{Z}{<A_1>}\right)^2 -  \left(\frac{Z}{<A_2>}\right)^2 \right]}.
 \label{eq:csym_exact}
\end{eqnarray}

The $<A>^{-n}$ ($n>0$) dependence of these quantities rend the 
corrections negligeable for heavy fragmentation products \cite{ono}
and quantitative results will be presented for a case relevant to most
multifragmentation reactions.

The approximations of
Eq. (\ref{eq:csym_ono}) and Eq. (\ref{eq:csym_tsang}) may be judged within a
microcanonical multifragmentation model \cite{micromodels}.
We use the MMM version \cite{mmm}
where the equilibrated multifragmenting source
is characterized by its mass $A_s$, charge $Z_s$,
excitation energy $E$, total momentum ${\rm P}$,
total angular momentum  ${\rm L}$ and freeze-out volume $V$.
All configurations
$C:\{\{ A_i, Z_i, \epsilon_i, {\rm p}_i, {\rm r}_i\}, \{ i=1,...,NF \} \}$
permitted by the specific microcanonical conservation laws are spanned
by a Metropolis trajectory and average values of physical quantities are calculated 
numerically on the basis of the statistical weight of each configuration $C$.
Break-up fragments are treated as normal nuclear density malleable objects
described by a zero-temperature binding energy as in Eq. (\ref{eq:bld})
where $a_v$=15.4941 MeV, $a_s$=17.9439 MeV,
$a_i$=1.7826, $a_c$=-0.7053 MeV and $a_a$=1.1530 MeV \cite{ld}.
This liquid drop description is consistent with a semiclassical Thomas-Fermi 
approximation \cite{ringschuck} or hot Hartree Fock \cite{vautherin}, 
where the effect of temperature is a modified occupation of the single 
particle eigenstates of the mean field hamiltonian. The finite temperature 
fragment energy functional in this approach is thus modified respect to the ground state
only for the internal excitation energy ($\epsilon$) coming from the occupation of 
continuum states. To avoid double counting of the free particles states \cite{bonche},
the internal energy is cut at the binding level.
%
%and are allowed to absorb internal energy ($\epsilon$)
%up to the binding level.
%We mention that the decoupled treatment
%of fragment formation and internal excitation
%results in different values of the kinetic and fragment internal temperatures,
%the second one being much smaller than the first one.
In our microcanonical formalism, the temperature is univocally defined
through the thermodynamic equality 
%In all formulae, $T$ stands for the temperature calculated according to
%the thermodynamical definition,
$T^{-1}=\partial S/\partial E=<(3N/2-4)/K>$, 
where $K$ represents the thermal kinetic energy, $N$ is the product multiplicity
and the last equality stems from the equipartition theorem applied to a non interacting
system of clusters \cite{rad99}. It is worthwhile mentioning that, due to the high energy
cut-off in the cluster level densities, the internal fragment excitation energy
cannot be used to estimate the temperature, and a thermometer based upon internal fragment
properties would severely underestimate the thermodynamic temperature.
In the experimental evaluation of $C_{sym}$ 
conversely, the temperature is also estimated from fragment properties or model calculations,
which may lead to an extra source of uncertainty. 

%Relevant for the purpose of the present paper is that, despite excited,
%the fragments are defined by a ground state binding energy. 
%This means that the value of $C_{sym}$ calculated out of isotopic yields has
%to be compared with the ground state symmetry energy.

The break-up stage of the decay which contains all information relevant for the
equation of state is completed with a particle evaporation stage which simulates
the subsequent disintegration of the excited fragments \cite{mmm}.
This step is important as well as it shows the capacity of experimentally
measurable quantities to access the physics at break-up.

The predictive power of
Eqs. (\ref{eq:csym_ono}) and (\ref{eq:csym_exact}) 
has been systematically checked
by confronting their results with the input symmetry energy which enters
fragment definition via the binding energy.
We have considered a variety of situations in which the size
of the emitting sources was varied between $A=230$ and $A=100$,
the freeze-out volume covered the usually accepted interval ($V=4V_0-8V_0$) and
the excitation energy ranged between 2 and 10 MeV/nucleon.
The lower limit of the source size for which
Eqs. (\ref{eq:csym_ono}), (\ref{eq:csym_exact}) and (\ref{eq:csym_tsang})
hold was conditioned by the possibility to calculate $\alpha$ out of the
isotopic composition of light emitted fragments.
Indeed, for small sources $\alpha$ manifests a relatively strong dependence on the
emitted cluster size, its calculation as an average value getting disputable 
\cite{microiso}. For completeness, the stability of the above mentioned formulas was
checked against sources isospin variation as well.
The conclusions are the same and only few illustrative cases will be considered
in the following.

\begin{figure}
\resizebox{0.6\textwidth}{!}{%
  \includegraphics{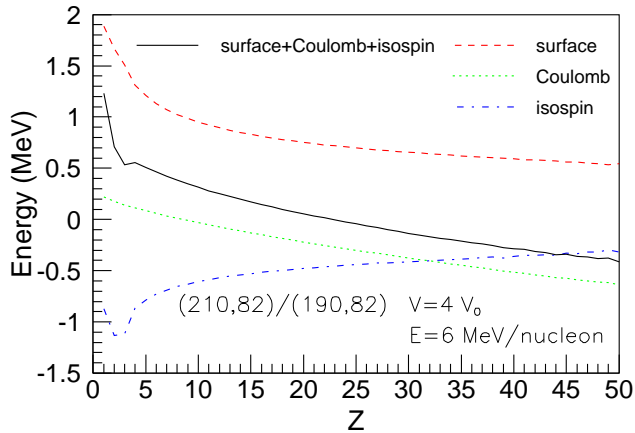}}
\caption{(Color online)
MMM predictions for the additional surface (dashed),
Coulomb (dotted) and isospin (dot and dash) contributions to the 
symmetry energy as a function of $Z$ at break-up.
The solid curve depicts the sum of the above mentioned terms.
The equilibrated systems are (210, 82) and (190, 82)
with $V=4V_0$ and excitation energy 6 MeV/nucleon.
}
\label{fig:corr}
\end{figure}

Fig. \ref{fig:corr} illustrates the magnitude of the additional terms of
Eq. (\ref{eq:csym_exact}) corresponding to corrections due to surface, Coulomb and
isospin energies as a function of the charge of the considered fragment
in a case typical for nuclear multifragmentation reactions, (210, 82) and (190,82) with
$V=4V_0$ ($V_0$ is the volume at normal nuclear density)
and $E=6$ MeV/nucleon. The values of the order of unity rend these corrections
negligeable even for light fragments, in agreement with expectations of Ref. \cite{ono}.
For this reason, the discussion on the possibility to extract the symmetry energy
from multifragmentation data will address only the original Eq. (\ref{eq:csym_ono}).

\begin{figure}
\resizebox{0.6\textwidth}{!}{%
  \includegraphics{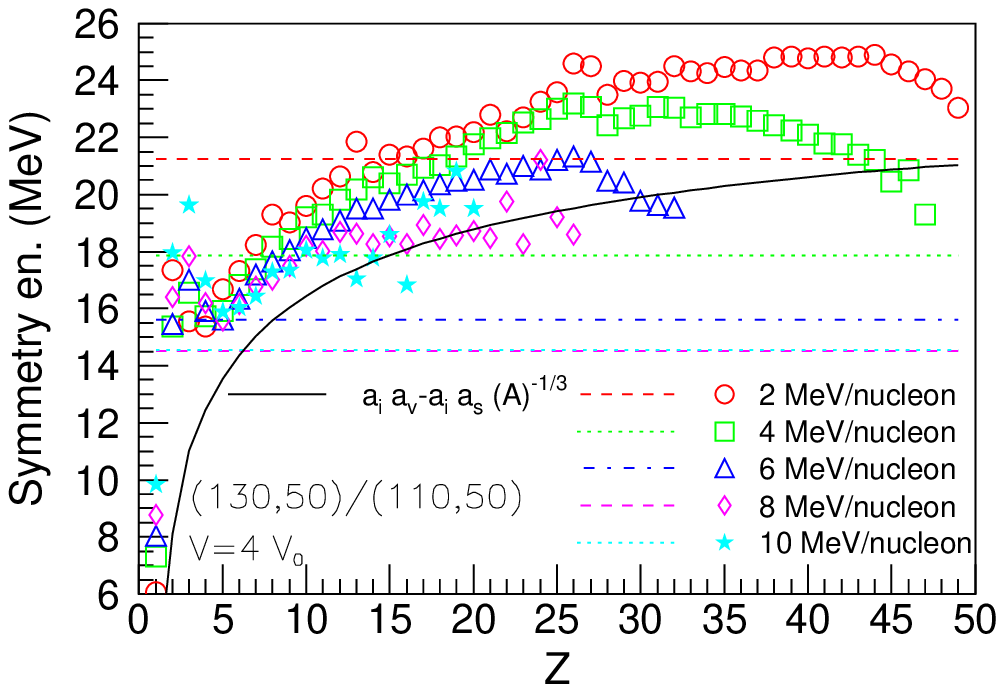}
  }
\resizebox{0.6\textwidth}{!}{%
  \includegraphics{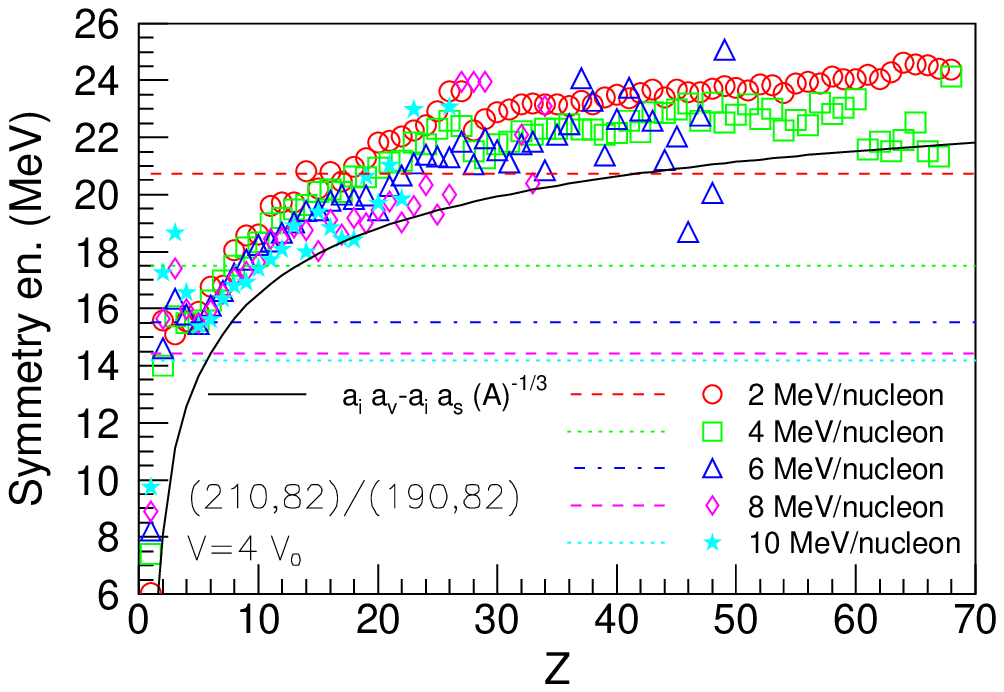}
  }
\caption{(Color online)
MMM predictions for the symmetry energy as a function of $Z$ at break-up.
The equilibrated systems are 
((130, 50) and (110, 50)) (upper panel) and
((210, 82) and (190, 82)) (lower panel)
with $V=4V_0$ and excitation energies ranging from 2 to 10 MeV/nucleon.
Solid curve: input symmetry energy of the model.
$C_{sym}$ calculated according to Eq. (\ref{eq:csym_ono}) is represented
with symbols while results of Eq. (\ref{eq:csym_tsang}) are illustrated
with dotted and dashed lines.}
\label{fig:csym_ex_z}
\end{figure}

Fig. \ref{fig:csym_ex_z} presents break-up stage predictions of MMM
for the asymmetry term as a function of the emitted fragment charge.  
Results of Eqs.  (\ref{eq:csym_ono}) and (\ref{eq:csym_tsang})
are shown for the multifragmenting nuclei 
((130, 50) and (110, 50)) (upper panel) and
((210, 82) and (190, 82)) (lower panel)
with $V=4V_0$ and excitation energy ranging from 2 to 10 MeV/nucleon.
The MMM results (symbols) are compared to the  
input symmetry energy of the model,
$a_i a_v-a_i a_s A^{-1/3}$ (solid curve).
As one may notice, Eq. (\ref{eq:csym_ono}) 
shows a remarkable overall stability against excitation energy variations
but its behavior while modifying the source size or excitation energy deserves
a more attentive investigation.
Firstly, Eq. (\ref{eq:csym_ono}) shows a systematic overestimation of the real value
by up to 4 MeV for the largest fragments.
With the increase of the source size and excitation energy, the overestimation
slightly diminishes so that, for the $Z$=82 sources and the highest considered
excitation energy, 10 MeV/nucleon, the calculated $C_{sym}$ exceeds by less than 1 MeV
the real value.
This is an interesting manifestation of the applicability conditions of grandcanonical approaches
in the case of small systems.
Thus, for relatively light fragments emitted by highly excited large systems,
grandcanonical formulas give reasonable values while more modest results are
obtained at low excitation energies and for fragments commensurable with the source size.
The bending of the curves calculated with Eq. (\ref{eq:csym_ono}) for the $Z$=50 sources at
2 and 4 MeV/nucleon is due to the mass and charge conservation specific of
a microcanonical ensemble.
Similar effect manifests also for the $Z$=80 case, but the limited charge domain
in the right panel of Fig. \ref{fig:csym_ex_z} hides it.
The residual slight overestimation of the input value of $C_{sym}$
in a domain where the grancanonical approximation should in principle 
be acceptable remains an open question and will be addressed in the future.
However, we have checked that it does not depend on the value of the high energy cutoff $\exp(-\epsilon/\tau)$
of the fragment internal density, as similar results are obtained when
$\tau \rightarrow \infty$. 
%and fragments internal and translational degrees
%of freedom are fully equilibrated.

%, may be due to the neglect of the fragment internal density 
%of states in Eq. (\ref{granca}), which because of the physical high energy cutoff
%violates the simple Boltzmann form Eq. (\ref{granca}).  

As expected, Eq. (\ref{eq:csym_tsang}) which approximates the isospin content
of fragments to the one of the source gives good results only for very heavy
fragments and low excitation energy. However in the fragmentation regime
it leads to a systematic underestimation of the symmetry energy. In particular,
Fig. \ref{fig:csym_ex_z} suggests that the low value $C_{sym}\approx$ 15 MeV extracted 
in some experimental analyses \cite{lefevre,frs} may be compatible with standard
ground state values for the symmetry energy. In this respect 
it is interesting to remark that the use of Eq. (\ref{eq:csym_tsang}) 
produces an unphysical apparent reduction of $C_{sym}$ with increasing excitation,
similar to the findings of Ref. \cite{lefevre}.
The explanation of this evolution in the model lays in the steep
diminish of $\alpha$ not compensated by the temperature increase \cite{microiso}.

\begin{figure}
\resizebox{0.6\textwidth}{!}{%
\includegraphics{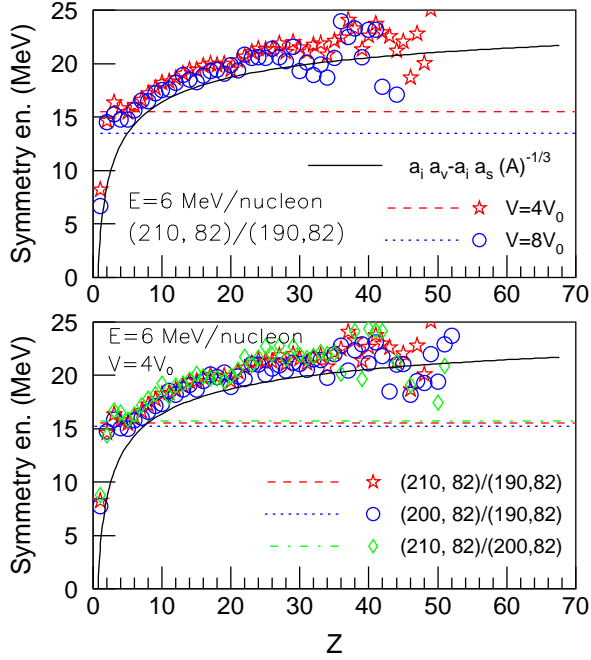}}  
\caption{(Color online)
MMM predictions for the symmetry energy as a function of $Z$ at break-up
as obtained using Eq. (\ref{eq:csym_ono}) (open symbols).
The stability against freeze-out volume variation ($V=4V_0$ and $V=8V_0$)
is illustrated for the equilibrated systems (190, 82) and (210, 82)
with 6 MeV/nucleon excitation energy (upper panel).
The stability against sources isospin modification is checked
for three pairs of $Z$=82 nuclei with $V=4V_0$ and 6 MeV/nucleon excitation energy
(lower panel).
Results of Eq. (\ref{eq:csym_tsang}) are indicated using dotted and dashed lines.}
\label{fig:csym_n_iso_z}
\end{figure}

Fig. \ref{fig:csym_ex_z} was obtained for a specific choice of the fragmenting
sources. However it was proved in Ref. \cite{microiso} that the isoscaling parameter $\alpha$
manifests a complex dependence on the considered pair of sources and their
equilibrium state under the microcanonical constraint.
This fact requires a detailed investigation
of the persistence of Eq. (\ref{eq:csym_ono})
under various freeze-out volumes, source isospin combinations and source sizes.
Therefore, Fig. \ref{fig:csym_n_iso_z} presents the results of
Eq. (\ref{eq:csym_ono}) for the same sources ((210, 82) and (190, 82))
with 6 MeV/nucleon excitation energy when the
freeze-out volume has different values ($V=4V_0$ and $V=8V_0$) (upper panels),
while the behavior with respect to sources isospin modification is represented
considering three pairs of $Z$=82 nuclei ((190,82), (200, 82) and
(210, 82))
at $V=4V_0$ and 6 MeV/nucleon excitation energy (lower panels).
As in the previous case, the solid curve illustrates the $(a_i a_v-a_i a_s A^{-1/3})$ term.
The conclusions confirm the stability of Eq. (\ref{eq:csym_ono}) while modifying
source isospin and freeze-out volume.
A more increased predictive power is expected
for larger sources with a more advanced fragmentation, where
the grandcanonical approximation is more reasonable.

\begin{figure}
\resizebox{0.6\textwidth}{!}{%
\includegraphics{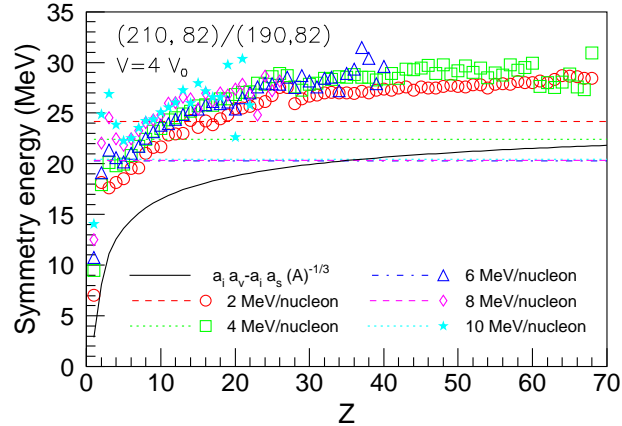}
  }  
\caption{(Color online)
MMM predictions for the symmetry energy as a function of $Z$
in the asymptotic stage of the decay as obtained using Eq. (\ref{eq:csym_ono})
(open symbols).
The equilibrated systems are (190, 82) and (210, 82)
with $V=4V_0$ and excitation energies ranging from 2 to 10 MeV/nucleon.
Results of Eq. (\ref{eq:csym_tsang}) are indicated using dotted and dashed lines.
In all cases, asymptotic values of $\alpha$ and break-up values of $Z/<A>$
have been used.}
\label{fig:csym_ex_ev_z=80}
\end{figure}

If the preceding figures suggest that 
for large systems and high excitation energies, 
Eqs. (\ref{eq:csym_ono}) and (\ref{eq:csym_exact})
may allow to extract the symmetry energy coefficient 
at the fragmentation stage from isoscaling observables, it is important to stress
that not only the average fragment isotopic composition, but also the isoscaling parameter
$\alpha$ should be known at the time of fragment formation, while the measured $\alpha$
value may have been distorted by sequential decay.
The effect of secondary decay on the isoscaling parameter is explored in Fig. \ref{fig:csym_ex_ev_z=80}.
Since in the MMM model the value of $\alpha$ is larger in the asymptotic stage of the reaction
than at break-up \cite{microiso}, Eqs. (\ref{eq:csym_ono}) and (\ref{eq:csym_exact})
lead to an overestimation of the symmetry energy coefficient.
If fractionation is neglected using Eq. (\ref{eq:csym_tsang}) these two errors tend
to compensate and the extracted values of  
$C_{sym}$ range from 20 to 25 MeV,
in better agreement with the real value
(if sufficiently heavy isotopes are analyzed for the surface influence to be negligible, 
which is not the case in present fragmentation data).
It is important to stress that this compensation entirely relies on the evolution
of $\alpha$ from hot to cold fragments.
Unfortunately, in this respect, information in the literature is controversial.
Thus, MSU-SMM \cite{msu-smm} and IQMD \cite{iqmd} plead in favor of negligible contribution
of sequential evaporation on $\alpha$ as stated in Refs. \cite{tsang} and \cite{iqmd_alpha},
respectively.
Stochastic mean field \cite{bob}, AMD \cite{ono} and
the Markov-chain SMM \cite{botvina-smm} with $C_{sym}>$15 MeV 
indicate $\alpha$ values decreasing from the break-up to the asymptotic stage 
\cite{bob_alpha,ono_comment,lefevre}.
Finally, MMM, EES \cite{ees} and
the Markov-chain SMM with $C_{sym}<$15 MeV
predict the opposite effect:
$\alpha$ values increasing from the break-up to the asymptotic stage
\cite{microiso,tsang,lefevre}.
The origin of these discrepancies could by due to isospin population
of break-up fragments, their excitation energy and secondary decay procedure as well, 
but a pertinent discussion of this issue goes much beyond the limits of present paper.

To summarize, we investigated 
the possibility of inferring the symmetry energy coefficient from isoscaling 
observables in multifragmentation reactions.
Our results indicate that, in the limit of large systems and high excitation energies
where the use of a grandcanonical approach is reasonable,
Eq. (\ref{eq:csym_ono}) is a reliable
tool to determine $C_{sym}$ out of the isotopic composition of the break-up fragments
but overestimations of few MeV are expected for the systems
usually produced in multifragmentation reactions.
In particular low values of $C_{sym}$ respect to the standard saturation density bulk value
may be interpreted as an effect of surface contributions to the symmetry energy.

Conversely, if we assume that the density of break-up fragments is close to their ground
state density \cite{micromodels}, the symmetry energy coefficient can be considered as known 
and isoscaling measurements can be used to probe fractionation. In particular spinodal 
decomposition is predicted to lead to higher fractionation than phase separation at equilibrium
\cite{baran}, and this effect could be accessed through isoscaling observables. 
To realize this ambitious program it is essential to experimentally control the effect 
of side feeding on the $\alpha$ isoscaling parameter. 
%In the particular case of MMM, the simplified expression which uses the source isotopic
%composition proves accurate enough 
%if heavy fragments ($Z\ge 30$) are isotopically resolved \cite{frs}.

\end{document}